\documentclass[a4paper]{jpconf}

\usepackage{amssymb}
\usepackage{amsfonts}
\usepackage{amsmath}
\usepackage{bm}
\usepackage{epsfig}
\usepackage{epsf}
\usepackage[]{graphicx}

\usepackage{color}
\usepackage{cite}
\usepackage{xspace}

\renewcommand {\d} {{\rm d}}

\newcommand {\E}  {{\varepsilon}}

\newcommand {\bfp} {{\bf p}}
\newcommand {\bfr} {{\bf r}}
\newcommand {\bfv} {{\bf v}}

\newcommand {\bfR} {{\bf R}}

\newcommand{\MBNExplorer}{\textsc{MBN Explorer}\xspace}

\newcommand{\Nacc}{{N_{\rm acc}}}
\newcommand{\Lch}{{L_{\rm ch}}}
\newcommand{\Nch}{{N_{\rm ch}}}

\begin{document}

\title{
Simulations of electron channeling in bent silicon crystal}

\author{G B Sushko$^{1}$, 
V G Bezchastnov$^{1}$\footnote{On leave from 
Department of Theoretical Astrophysics, 
Ioffe Physical-Technical Institute, Politechnicheskaya Str. 26, 
194021 St. Petersburg, Russia
and from St. Petersburg State Polytechnical Universty,
Politechnicheskaya  29, 195251 St. Petersburg, Russia},
A V Korol$^{1,2}$,
Walter Greiner$^{1}$,
A V Solov'yov$^{1}$
\footnote{On leave from A.F. Ioffe Physical-Technical Institute, St. Petersburg, Russia},
R G Polozkov$^{3}$,
V K Ivanov$^{3}$}

\address{$^1$ Frankfurt Institute for Advanced Studies, 
Ruth-Moufang-Str. 1, 60438 Frankfurt am Main, Germany}
\address{$^2$ St. Petersburg State Maritime University, 
Leninsky ave. 101, 198262 St. Petersburg, Russia}
\address{$^3$ St. Petersburg State Polytechnic University, 
Politekhnicheskaya str. 29, 195251 St. Petersburg, Russia}

\ead{korol@fias.uni-frankfurt.de}

%\pacs{02.70.Ns, 02.70.Uu, 41.60.-m, 61.85.+p}

%%%%%%%%%%%%%%%%%%%%%%%%%%%%%%%%%%%%%%%%%%%%%%%%%%%%%%%%%%%%%%%%%
\begin{abstract}
We report on the results of theoretical simulations of the electron channeling 
in a bent silicon crystal. 
The dynamics of ultra-relativistic electrons 
in the crystal is computed using the newly developed part \cite{NewPaper_2013} of 
the MBN  Explorer package \cite{MBN_ExplorerPaper,MBN_ExplorerSite},
which simulates classical trajectories of in a crystalline medium by integrating
the relativistic equations of motion with account for the interaction between the 
projectile and crystal atoms.
A Monte Carlo approach is employed to sample the incoming electrons and to 
account for thermal vibrations of the crystal atoms.
The electron channeling along Si(110) crystallographic planes are studied for the 
projectile energies 195--855 MeV and different curvatures of the bent crystal.
\end{abstract}

%%%%%%%%%%%%%%%%%%%%%%%
\section{Introduction}

Channeling is a phenomenon which is known to arise when a charged particle 
enter a crystal at angles small enough with respect to crystallographic planes 
or axes \cite{Lindhard}. 
The particles become confined and forced to move through the 
crystal preferably along the corresponding crystallographic direction
experiencing the collective action of electrostatic field of the 
lattice ions.
The field is repulsive for positively charged particles and, therefore, 
they are steered into the inter-atomic region, while negatively charged 
projectiles move in the close vicinity of ion strings or planes.

Under certain conditions \cite{Tsyganov1976}
the channeling effect persists even if a crystal is bent.
In this case, the particle deviates from its initial direction 
of motion due to extremely strong electrostatic field in the crystal.
The field strength is typically of the order of $10^{10}$ V/cm which 
is equivalent to the magnetic field of approximately 3000 T.
Therefore, bent crystal can steer particles much more effectively than 
any existing magnet.
In particular, in experimental high energy physics the crystals with 
bent crystallographic planes are used to steer high-energy charged particle 
beams replacing huge dipole magnets (see, e.g. \cite{Uggerhoj2011}).

A comprehensive review of theoretical and experimental 
achievements in the investigation of the channeling effect in straight and 
bent crystals one can find in \cite{Lindhard,Gemmel,Sorensen1996,%
BiryukovChesnokovKotovBook,Uggerhoj_RPM2005,Uggerhoj2011}.

Recently, the crystalline undulator (CU) concept was formulated 
for producing undulator-like electromagnetic radiation in the sub hundreds of 
keV up to the MeV photon energy range \cite{KSG1998,KSG_review_1999}.
In a CU the radiation is emitted by a beam of ultra-relativistic charged particles 
which undergo channeling in a periodically bent crystal.  
As a result, in addition to a well-known channeling radiation, there appears 
the radiation due to the undulating motion of the particles which follow the periodic 
bending of crystallographic planes. 
The intensity and characteristic frequencies of the CU radiation can 
be varied by changing the type of channeling particles,
the beam energy, the crystal type  and the parameters of periodic bending.
We refer to Ref. \cite{ChannelingBook2013} which represents 
the underlying fundamental physical ideas as well as the theoretical, 
experimental and technological advances made during the fifteen years
in exploring various features of CUs and its radiation.

Several experimental attempts were made \cite{BaranovEtAl_CU_2006} 
or planned to be made \cite{Backe_EtAl_2011a} 
to detect the radiation from a positron-based CU.
So far, the attempts have not been successful due to
various reasons (see discussion in Refs. 
\cite{Backe_EtAl_2011a,ChannelingBook2013}).
However, quite recently the first signatures showing evidence for the 
CU radiation were experimentally observed for 195--855 MeV electrons 
at the Mainz Microtron (Germany) facility
\cite{Backe_EtAl_2011,BackeLauth_2012}.
Another set of experiments is scheduled for the year 
2013 at the SLAC facility (USA) with 10 -- 20 GeV electron 
beam \cite{Uggerhoj_2012}.

Theoretical support of ongoing and future experiments as well as accumulation
of numerical data on channeling and radiative processes of ultra-relativistic 
projectiles in straight, bent and periodically bent crystals must be based on an 
accurate procedure which allows one to simulate the trajectories 
corresponding to the channeling and non-channeling regimes.
The procedure must include a rigorous description of the particle motion   
and an efficient algorithm of its numerical realization.
It is strongly desirable to make the procedure as much model-independent as possible.
This means that the only allowed parameters are 
those which describe {\em pairwise} interactions of the 
projectile with constituent atoms. 

Most of the existing codes, capable to simulate channeling process, 
can be categorized as following.
Some of them (see, for example, Refs. \cite{Artru1990,Biryukov1995,Maisheev1996})
are based on the concept of the continuous potential \cite{Lindhard}.
This approximation, being adequate in describing the channeling motion, becomes
less accurate and more model-dependent when accounting for uncorrelated 
scattering events. 
The accurate description of the latter is essential for a quantitative analysis
of the dechanneling and rechanneling processes.
Other group of the codes 
\cite{BakEtal1985,SmuldersBoerma1987,ShulgaSyshchenko2005} 
utilizes the scheme of binary collisions which
assumes that the motion of a projectile at all times is influenced by the 
force due to the nearest atom.
Computer facilities available at present allow one to go beyond this limitation
and to account for the interaction with larger number of the crystal atoms.
Such an extension of the binary collisions algorithm was implemented in the 
recent code for electron channeling described in 
Refs. \cite{KKSG_simulation_straight}.
The code, however, was based on the specific model for electron--atom scattering
which results in a noticeable overestimation of the mean scattering angle.
In more detail this topic was addressed in Ref. \cite{NewPaper_2013} (see also 
comments in Section \ref{NumericalResults}). 

To simulate propagation of ultra-relativistic
particles through media, the channeling process in particular, 
one can  approaches and algorithms used in modern 
molecular dynamics codes (see Ref. \cite{MBN_ExplorerPaper} for the 
comparative review of the codes). 
The latter allow one to model the dynamics of various molecular system by 
efficient numerical integration of classical equations of motion for all 
atoms in the system.
The interaction between atoms is implemented in terms of interatomic
potentials, the types and parameters of which can be chosen from a broad 
range to ensure the most adequate quantitative description of the simulated 
molecular system.

Recently, the molecular dynamics concept was applied to describe the motion of 
a single projectile in the static field of atoms which constitute a scattering 
medium \cite{NewPaper_2013}. 
For this purpose we have used the \MBNExplorer software
package \cite{MBN_ExplorerPaper,MBN_ExplorerSite} and supplied it 
with additional module which treats classical relativistic equations of motion 
and generates the crystalline
environment dynamically in the course of particle propagation.
This module, combined with the variety of interatomic potentials implemented
in \MBNExplorer, makes the program a unique tool for studying relativistic
phenomena in various environments, such as crystals, amorphous bodies,
biological medium.
Verification of the code against available experimental data 
was carried out for $6.7$ GeV projectile electrons and positrons in 
channeling in straight Si(110) channels 
\cite{NewPaper_2013}.

In the present paper we present the results obtained with the new code \cite{NewPaper_2013}
on the Monte Carlo simulation of the channeling of electrons 
along the (110) crystallographic plane in straight and bent single silicon crystal. 
The calculations were performed for $\E=195, 345, 600$ and 855 MeV electrons 
which correspond to the experimental conditions at the 
Mainz Microtron (Germany) \cite{Backe_EtAl_2008}.  
We have restricted the current analysis to the bent crystals 
with constant curvature, although these results will help one 
to estimate the reasonable parameters of periodically bent crystal 
channels and, therefore, will facilitate future simulations and 
experimental study of the electron-based crystalline undulator.

%%%%%%%%%%%%%%%%%%%%%%%%%%%%%%%%%%%%%%%%%%
\section{Physical model for channeling \label{Model}}

Upon entering a crystal, a particle with the charge $q$ experiences an 
electrostatic field $U({\bf r})$ of the crystalline environment. 
To compute the motion of projectiles in the latter potential, we employ the 
relativistic classical dynamics and solve the following pair of equations 
\begin{equation}
\d\bfr/\d t = \bfv, 
\quad 
\d\bfp/\d t = -q\,\partial{U}/\partial{\bfr}
\label{dyn_eqs}
\end{equation}
for the particle coordinate $\bfr(t)$ and velocity $\bfv(t)$ as the functions of 
propagation time $t$. 
The momentum ${\bf p}$ relates to the velocity ${\bf v}$ 
according to the relativistic expression $\bfp = m\gamma\bfv$, 
where $m$ is the mass of the particle, 
$\gamma = [1-(v/c)^2]^{-1/2} = \E/(mc^2)$ is the Lorenz factor, $\E$ is 
the projectile energy and $c$ is the speed of light. 

To incorporate the many-body interaction of the projectile with the crystalline 
constituents, the potential $U$ is represented by the sum of atomic potentials 
$U_{\rm at}$, 
\begin{equation}
U(\bfr) = \sum_j U_{\rm at}\left(\bfr-\bfR_j\right),
\label{many-body}
\end{equation}
where $\bfR_j$ is the position vector of the $j$-th atom in the crystal. 
The atomic potentials are evaluated according to the Moli\`{e}re 
approximation \cite{Moliere}. % $[$40$]$. 
The computations with \MBNExplorer  can also be performed, 
if demanded, employing a more elaborated Pacios approximation 
for the atomic potentials \cite{Pacios1993}. %$[$41$]$. 

Formally, the sum in Eq.~(\ref{many-body}) is carried out over all atoms of the 
sample. 
However, the atomic potentials decay with increasing distance from the atoms, and 
the decay becomes exponential when the distances exceed the atomic sizes. 
This allows one to restrict the number of atoms interacting with the projectile 
by accounting only for those separated from the projectile by distances shorter 
than some cut-off distance. 
Selection of these atoms for a given location of the projectile is facilitated by a 
linked cells algorithm implemented in  \MBNExplorer 
\cite{MBN_ExplorerPaper,MBN_ExplorerSite}. 

The module for channeling simulations, implemented in \MBNExplorer, 
has been designed to describe the crystalline structure acting on a moving projectile 
"on the fly" in the course of integrating the dynamical equations~(\ref{dyn_eqs}). 
This is achieved by introducing a "dynamic simulation box" which shifts following 
the propagation of the particle (see Ref. \cite{NewPaper_2013} for more details). 
Within the simulation box, the equilibrium positions of the crystal nodes are 
determined according to the lattice structure under study. 
Then, the thermal vibrations of the crystal are accounted for by displacing the 
atoms with respect to the equilibrium positions. 
For each atom, a displacement $\Delta_i$ along a given 
Cartesian axis $i=x,y,z$ is selected randomly according to the normal distribution 
\begin{equation}
w(\Delta_i) = \left(2\pi{u_T^2}\right)^{-1/2}\exp\left[-\Delta_i^2/2u^2_T\right],
\label{therm_fluct}
\end{equation}
where $u_T$ is the amplitude of the thermal vibrations. 
For a Silicon crystal at the room temperature $u_T$ = 0.075 \AA
\cite{Gemmel}. % $[$42$]$.

To study the channeling motion along a particular crystallographic 
plane one needs to specify the orientation of the crystal with respect to the 
incoming particles. 
To do this, we introduce the laboratory reference frame with the $z$-axis along the 
beam.
In the case of a straight crystal we orient the crystallographic plane parallel 
to the $(xz)$-plane. 
A care is taken to avoid accidental orientations of 
crystallographic axes along the beam, as this is also 
done in the most common experiments in order to exclude the axial channeling. 
To simulate a bent crystal, the coordinates 
$x',y',z'$ of each lattice node are obtained from the coordinates 
$x,y,z$ of the same node in the straight crystal according to the relations  
$x' = x$, $y' = y + \delta(z)$, $z' = z$, where
\begin{equation}
\delta(z) = R - \sqrt{R^2-z^2} 
\label{bending}
\end{equation}
is the shape of the bent crystallographic plane and $R$ is the bending radius. 

A convenient parameter showing impact of the curvature $R^{-1}$ on the channeling 
for the particles with the energy $\E$ is the ratio $C$ of two opposite 
forces acting on the projectile: the centrifugal force $\E/R$ and a maximal 
interplanar force $F_{\rm max}$ stabilizing the motion of the 
particle transversely to the crystallographic planes. 
The value $C = 0$ corresponds to the straight crystal 
and $C = 1$ is the critical value at which the potential barrier between the 
channels disappear. 
The value $F_{\rm max}$ has been estimated withing the 
continuous approximation for the planar potentials and found to be 
$6.67$~GeV/cm for the electron channeling along the 110 plane in Silicon
crystal. 
We use the latter value as the reference to introduce the bending parameter
\begin{equation}
C = 1.57\times{10^{-4}}(\E_{\rm MeV}/R_{\rm cm}),
\label{C}
\end{equation}
where $\E_{\rm MeV}$ is the electron energy in MeV and $R_{\rm cm}$ is the bending 
radius in cm.

Having specified the orientation and bending of the crystal with respect to the 
beam direction, we use a Monte Carlo approach to sample the beam electrons. 
The particles are assumed to enter the crystal at $t=0$ with $z=0$ and the 
aperture $(x,y)$ coordinates being randomly selected from the intervals 
$-1.5\,d \leq x \leq 1.5\,d$ and $-0.5\,d \leq y \leq 0.5\,d$, 
where $d$ is the interplanar distance  
(e.g., $d = 1.92$~{\AA} for the Si (110) planes). 
The corresponding initial velocities are set to be $v_x = v_y = 0$ 
which correspond to the ideal case of a zero-emittance beam.
Thus, at the entrance the particles are directed along the crystalline planes, 
and the particular sampling of the initial $y$-coordinates covered uniformly 
the interplanar extension for the channel which can guide the particles into the 
crystal. 
With the above initial conditions, the dynamical equations~(\ref{dyn_eqs}) are 
integrated  numerically employing the forth-order Runge-Kutta algorithm. 
The integration proceeds until $z$ exceeds the length $L$ of the crystalline sample. 

%%%%%%%%%%%%%%%%%%%%%%%%%%%%%%
%\section{Channeling parameters}
\section{Numerical results \label{NumericalResults}}

Using the approach described in the preceding section we have simulated propagation 
of the electrons and positrons with energies $\E = 195$, $345$, $600$ and $855$ MeV along 
the (110) planes in straight and bent single crystal of the length $L$ from 25 up to 300 
micron. 
Most of the simulations have been performed for $L = 75$ $\mu$m.

The Monte Carlo approach to sampling the incoming particles and including the 
thermal fluctuations of the crystalline lattice yields a manifold of different 
trajectories. 
By analyzing the dependencies $y(z)$ for each trajectory one 
concludes whether the channeling occurs and, if so, what parts of this trajectory 
are in the channeling mode. 
The analysis is similar to that used for the  straight crystal \cite{NewPaper_2013},
whereas now the shape of bent channels is described by Eq.~(\ref{bending}). 

Although the particles are initially directed along the 
crystalline planes, not all of the trajectories show the channeling. 
Thus, the important parameter to estimate is the acceptance 
\begin{equation}
A = \Nacc/N_0, 
\label{acceptance}
\end{equation}
which is the ratio of the number $\Nacc$, of particles captures 
into the channeling mode at the crystal entrance (the accepted particles) to
the total number $N_0$ of the simulated trajectories 
(the incident particles). 
In general, the acceptance depends on the angular distribution of the particles 
at the entrance, in particular it decreases drastically for the incident angles 
exceeding the Lindhard's critical angle. 
In our simulations the incident beam 
has zero emittance, i.e. the projectiles are collimated "optimally" to get into the 
inside of the crystal. 
Therefore, the acceptance we calculate represents an upper 
bound estimate for a possible experimental outcome. 

\begin{figure}[h]
\centering
\includegraphics[scale=0.4,clip]{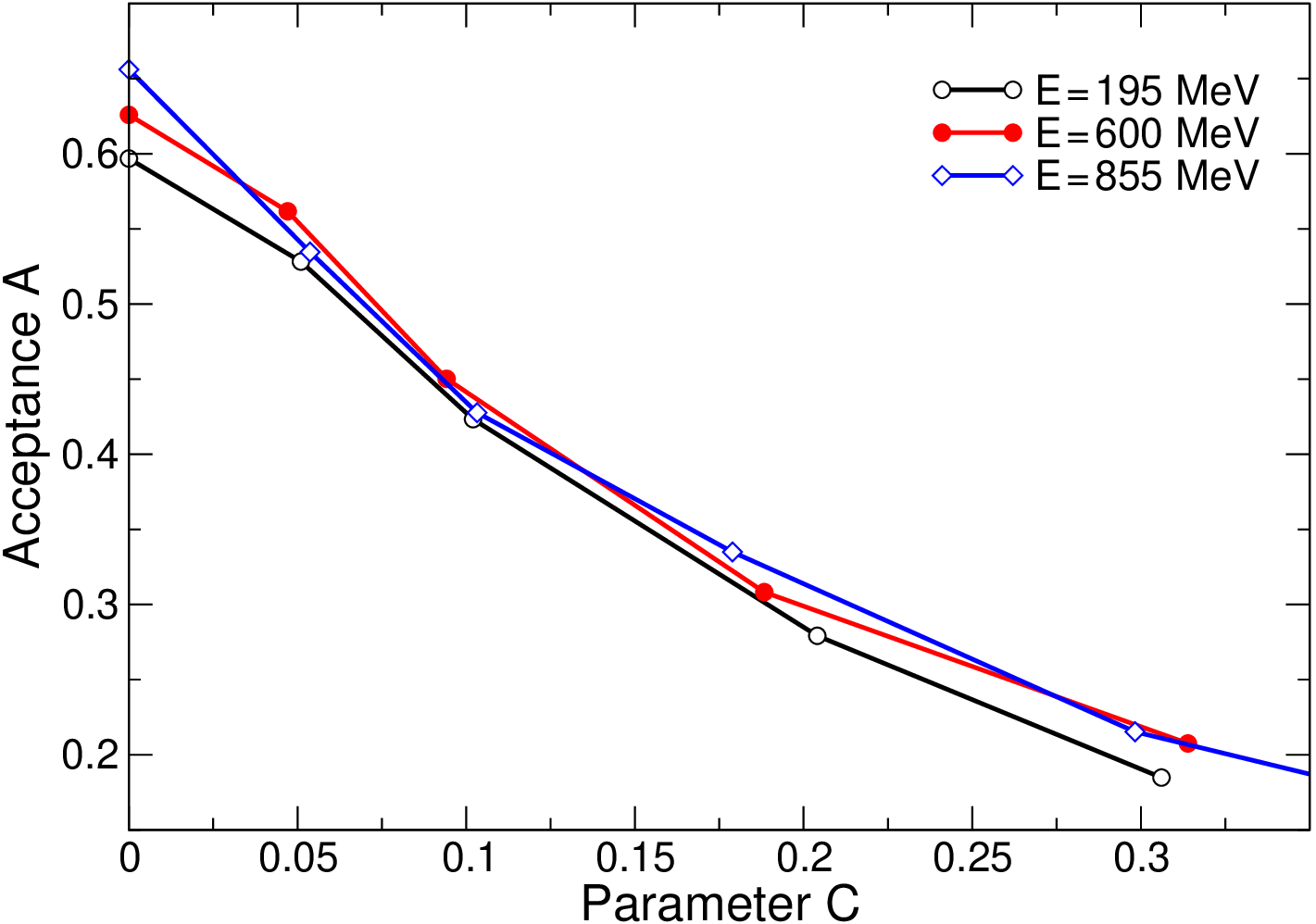}
\caption{Acceptance $A$ of the planar Si (110) channel 
as a function of the bending parameter $C$ for 
several energies of electrons as indicated.} 
\label{Acceptance}
\end{figure}

Fig.~\ref{Acceptance} shows the acceptance of the electrons of different energies
directed into the Si(110) channels.  
It is seen that for the same values of the bending parameter $C$ the acceptance
does not vary noticeably with the electron energy. 
Notice that the bending parameter is energy dependent itself, and the same 
values of this parameter correspond to larger bending radius with increasing energy 
(see Eq.~(\ref{C}). 
For each electron energy studied in Fig.~\ref{Acceptance}, the acceptance is maximal for 
the straight channel ($C=0$) and 
decreases with $C$ due to increasing centrifugal force acting on the projectile 
at the entrance. 

%%%%%%%%%%%%%%%%%%% Figure 1: Acceptance

To describe the channeling properties quantitatively, we have computed the 
numbers $N_{\rm ch0}(z)$ and $N_{\rm ch}(z)$ of the electrons that are in the
channeling mode at distance $z$ from the crystal entrance.
The former quantity stands for the number of electrons that propagate 
to $z$ channeling in the channel where they were accepted at the entrance. 
The quantity $N_{\rm ch}(z)$ is the number of electrons that 
are in the channeling mode inside the crystal, irrespective of the channel which 
guides the electronic motion at given $z$. 
With increasing $z$ the number $N_{\rm ch0}(z)$ decreases as the accepted electrons 
leave the entrance channel, i.e. a {\em dechanneling} takes place. 
In the contrast, the number $N_{\rm ch}(z)$ 
can increase when the electrons, including those not accepted at the entrance, 
can become channeling in some other channel as a result of {\em rechanneling}. 
The latter process results from random collisions with the crystal 
constituents experienced by the electrons that have left the channeling regime. 
The collisions can occasionally scatter the projectiles into the crystalline channels. 
If the scattering is accompanied by a reduction in the transverse 
energy, the projectile can become re-channeled. 

By normalizing $N_{\rm ch0}(z)$ with respect to the number $N_0$ of all simulated 
simulated and to the number $\Nacc$ of accepted electrons we determine the 
channeling fractions "1" and "2" which characterize the dechanneling process. 
We also introduce the channeling fraction "3" as the ratio of 
$N_{\rm ch}(z)$ to the number of accepted electrons. 
The latter fraction is determined by both dechanneling and rechanneling that 
happen to propagating electrons. 

\begin{figure}[h]
\centering
\includegraphics[scale=0.42,clip]{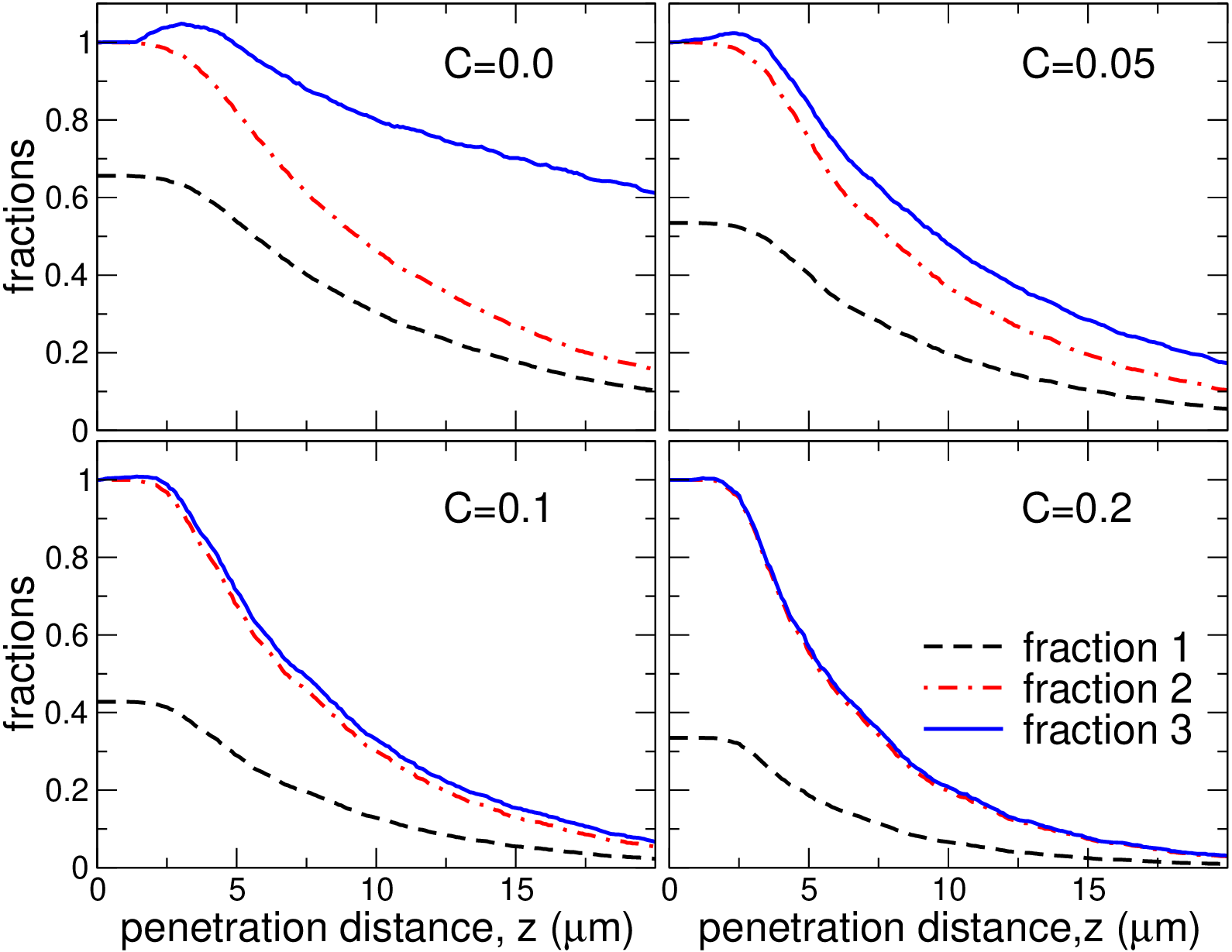}
\caption{Fractions of channeling electrons with energy 855 MeV for 
different values of the bending parameter $C$ as indicated. 
Fractions "1", "2" and "3" designate the ratios 
$N_{\rm ch0}(z)/N_0$, $N_{\rm ch0}(z)/N_{\rm acc}$ and $N_{\rm ch}(z)/N_{\rm acc}$, 
respectively.}
\label{fractions123}
\end{figure}

Fig.~\ref{fractions123} shows the fractions defined above for the electrons 
with energy 855~MeV channeling in Si(110). 
Four separate plots in the figure correspond to different values of 
the bending parameter. 
Fraction "1" differs from fraction "2" by a constant factor - the acceptance $A$, 
and is included in the plots with an illustrative purpose only. 
For the same propagation distance $z$, all three fractions become progressively 
smaller with increasing the bending parameter. 
More interesting is to compare fractions "2" and "3". 
Here, a large enhancement of fraction "3" with respect to "2" is seen for the 
straight ($C=0$) channel as well for bent channels corresponding to small values of the 
bending parameter, $C\lesssim 0.05$. 
In particular, for the straight channel at the distance $z=10$~microns 
the number of electrons channeling in {\em any channel} exceeds the number of 
electrons staying in the {\em entrance channel} by the factor of two, as a 
result of the dechanneling process. 
With $C$ increasing, the events of dechanneling become more rare, and the 
difference between the fractions diminishes (see the plot for $C=0.2$). 

%%%%%%%%%%%%%%%%%%% Figure 2: Fractions 1,2,3

\begin{figure}[h]
\centering
\centering
\includegraphics[scale=0.3,clip]{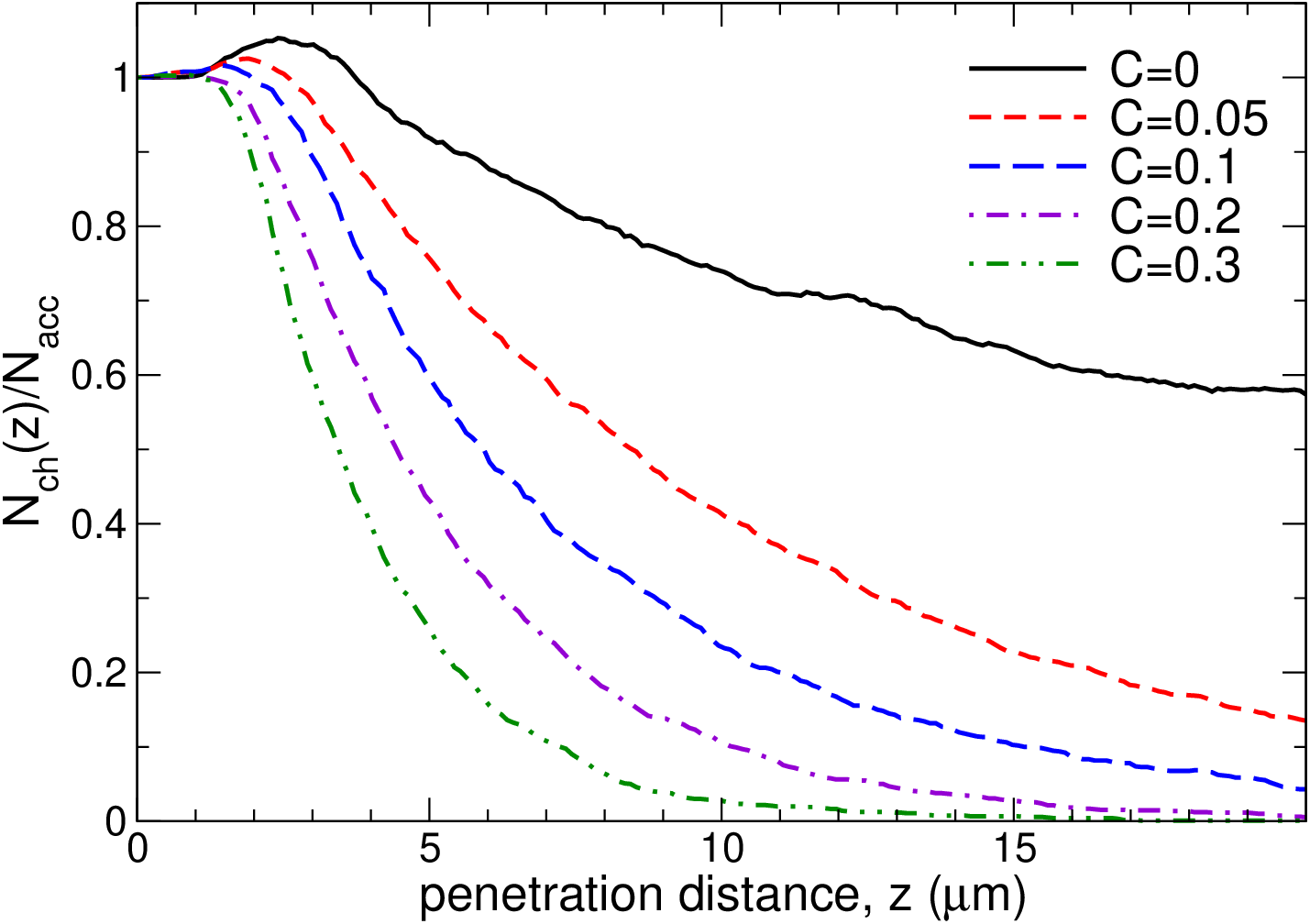}
\includegraphics[scale=0.3,clip]{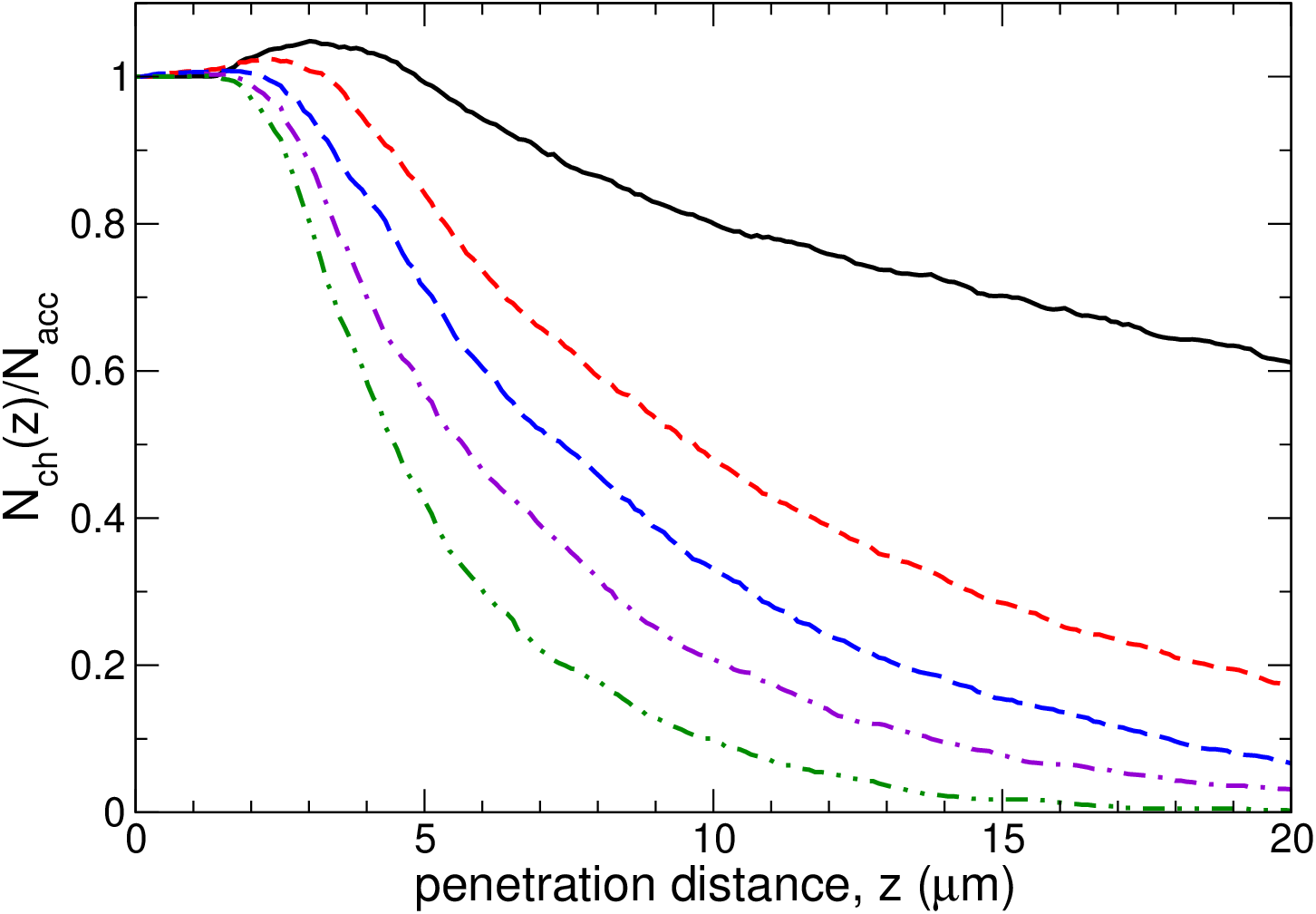}
\caption{The channeling fraction $N_{\rm ch}(z)/N_{\rm acc}$  
for  $\E=600$~MeV (left plot) and $\E=800$~MeV (right plot) electrons
versus the penetration distance. 
Different curves correspond to 
different values of the bending parameter as indicated. 
}
\label{fractions3}
\end{figure}

Fig.~\ref{fractions3} shows the fractions $N_{\rm ch}(z)/N_{\rm acc}$ of 
channeling electrons with the energies $\E=600$ and $855$~MeV for different values 
of the bending parameter $C$. 
Comparing the fractions for same values of $C$ one notices , that that the portion of 
channeling electrons is higher for higher $\E$. 
With $C$ increasing the channeling fractions gradually decrease and change 
the character of the decay at large distances: from the inverse power-law decay
at small $C$ to the exponential one at the larger values. 

%%%%%%%%%%%%%%%%%%% Figure 2: Fractions 1,2,3

By analyzing the channeling segments for each simulated trajectory we have 
calculate two characteristic penetration lengths introduced in Ref. \cite{NewPaper_2013}. 
The first one, $L_{\rm p1}$, is a mean value of the primary channeling segments 
which start at the crystal entrance. 
The other length, $L_{\rm p2}$, is defined as a mean path with respect to 
all channeling segments, including those which result due to the rechanneling 
process. 
Generally speaking, $L_{\rm p1}$ depends on the angular distribution of 
the particles at the entrance.
Our simulations, performed for of the zero emittance beams, yield an upper bound 
estimate for this length. 
However, the rechanneling electrons are statistically distributed over the incident 
angles that do not exceed the Lindhard critical angle. 
Therefore, the length $L_{\rm p2}$ can be thought for 
quantifying the penetration depth for a beam with non-zero emittance. 

%%%%%%%%%%%%%%%%%%%%%%%%%%
\Table{\label{Table_lengths}
Acceptance $A$ and penetration lengths $L_{\rm p1}$ and $L_{\rm p2}$ 
for different energies $\E$ of electrons channeling in bent Si(110). 
The results refer to the crystalline sample 
of the length $L=75$~$\mu$m.}
 \br
$\E$(MeV) &$R$ (cm) & $C$   &$N_0$ & $A$   &$L_{\rm p1}$ ($\mu$m)& $L_{\rm p2}$ ($\mu$m) \\ 
\mr
195       &$\infty$ & 0     & 1708 & 0.583 & $\04.08 \pm 0.29$   & $\04.08 \pm 0.11$  \\ 
          & 3       & 0.010 & 1570 & 0.575 & $\03.90 \pm 0.30$   & $\04.08 \pm 0.14$  \\ 
          & 0.6     & 0.051 & 1710 & 0.525 & $\03.91 \pm 0.29$   & $\03.91 \pm 0.23$  \\ 
          & 0.3     & 0.102 & 1550 & 0.423 & $\03.19 \pm 0.25$   & $\03.21 \pm 0.24$  \\ 
          & 0.1     & 0.306 & 1728 & 0.196 & $\01.99 \pm 0.22$   & $\01.99 \pm 0.21$  \\ 
\mr
600       &$\infty$ & 0     & 1420 & 0.609 & $\09.38 \pm 0.75$   & $\08.70 \pm 0.33$  \\ 
          & 2       & 0.047 & 1502 & 0.560 & $\08.81 \pm 0.72$   & $\08.25 \pm 0.62$  \\ 
          & 1       & 0.095 & 1461 & 0.446 & $\07.05 \pm 0.63$   & $\07.12 \pm 0.60$  \\ 
          & 0.5     & 0.188 & 1432 & 0.330 & $\05.28 \pm 0.52$   & $\05.28 \pm 0.52$  \\ 
\mr
855       &$\infty$ & 0     & 1128 & 0.656 & $11.90 \pm 1.11$    & $11.13 \pm 0.53$ \\ 
          & 2.5     & 0.054 & 1462 & 0.553 & $\09.79 \pm 0.81$   & $\09.87 \pm 0.72$  \\ 
          & 1.3     & 0.103 & 1419 & 0.423 & $\08.55 \pm 0.81$   & $\08.50 \pm 0.78$  \\ 
          & 0.7     & 0.192 & 1460 & 0.303 & $\06.59 \pm 0.68$   & $\06.61 \pm 0.68$  \\ 
          & 0.5     & 0.268 & 1504 & 0.241 & $\05.38 \pm 0.55$   & $\05.43 \pm 0.56$  \\ 
          & 0.3     & 0.447 & 1376 & 0.134 & $\04.50 \pm 0.61$   & $\04.48 \pm 0.61$  \\ 
 \br
 \end{tabular}
 \end{indented}
 \end{table}

Either one from $L_{\rm p1}$ and $L_{\rm p2}$ provides an estimate of 
the {\em dechanneling length}. 
Our results on these quantities for different electron energies and bending parameters are 
presented in Table~\ref{Table_lengths}. 
Within the statistical uncertainties, the values for both lengths yield the same estimates 
for the dechanneling length. 
Also included in this table are the numbers $N_0$ of the simulated 
trajectories along with the corresponding acceptances $A$. 

It is worth noting that the values of dechanneling lengths reported in the table 
noticeably exceed those estimated earlier for $\E=855$ MeV electrons channeled
in straight \cite{KKSG_simulation_straight,KKSG_NuovoCimento} and bent 
\cite{BentSi110_2011} Si (110) channels. 
The calculations performed in the cited papers were based
on the peculiar model of the elastic scattering of an ultra-relativistic projectile
from the crystal constituents.
The model assumes that due to the high speed of the projectile, its
interaction interval with a crystal atom is short enough
to substitute the atom with its ``snapshot'' image:
instead of the continuously distributed electron charge
the atomic electrons are treated as point-like charges placed at fixed
positions around the nucleus.
Next, the model implies that the interaction of an ultra-relativistic
projectile (e.g., an electron) with each atomic constituent can be reduced
to the classical Rutherford scattering.
In Ref. \cite{NewPaper_2013} it was demonstrated, that such
a ``snapshot'' model noticeably overestimates the mean scattering angle
in the process of elastic scattering in a single electron-atom collision.
Qualitatively, it is clear that substituting a ``soft'' electron cloud
with a set of point-like static electrons must lead to the increase of
the scattering angle simply because each electron acts as a charged
scatterer of an infinite mass.
As a result, the projectile experiences, on average, harder collisions
with electrons as compared to the case when they are continuously
distributed in space.
The mean square angle for a single scattering is a very important quantity
in the multiple-scattering region,
where there is a large succession of small-angle deflections symmetrically
distributed about the incident direction.
In particular, the mean square angle due to soft collisions defines the diffusion 
coefficient which, in turn, is proportional to the dechanneling length 
(see, for example, Refs. \cite{BiryukovChesnokovKotovBook,Backe_EtAl_2008}).
It was noted in Ref. \cite{NewPaper_2013} that the ``snapshot'' approximation
underestimates the dechanneling length of 855 MeV electrons in straight Si (110) 
by approximately 30 per cent.
Comparing the $L_{\rm p1}$ and $L_{\rm p2}$ values presented in Table \ref{Table_lengths} 
for $C>0$ with the data from Ref. \cite{BentSi110_2011} one concludes that the 
discrepancy becomes more pronounced for larger values of the bending parameter.

\begin{figure}[h]
\centering
\includegraphics[scale=0.42, clip]{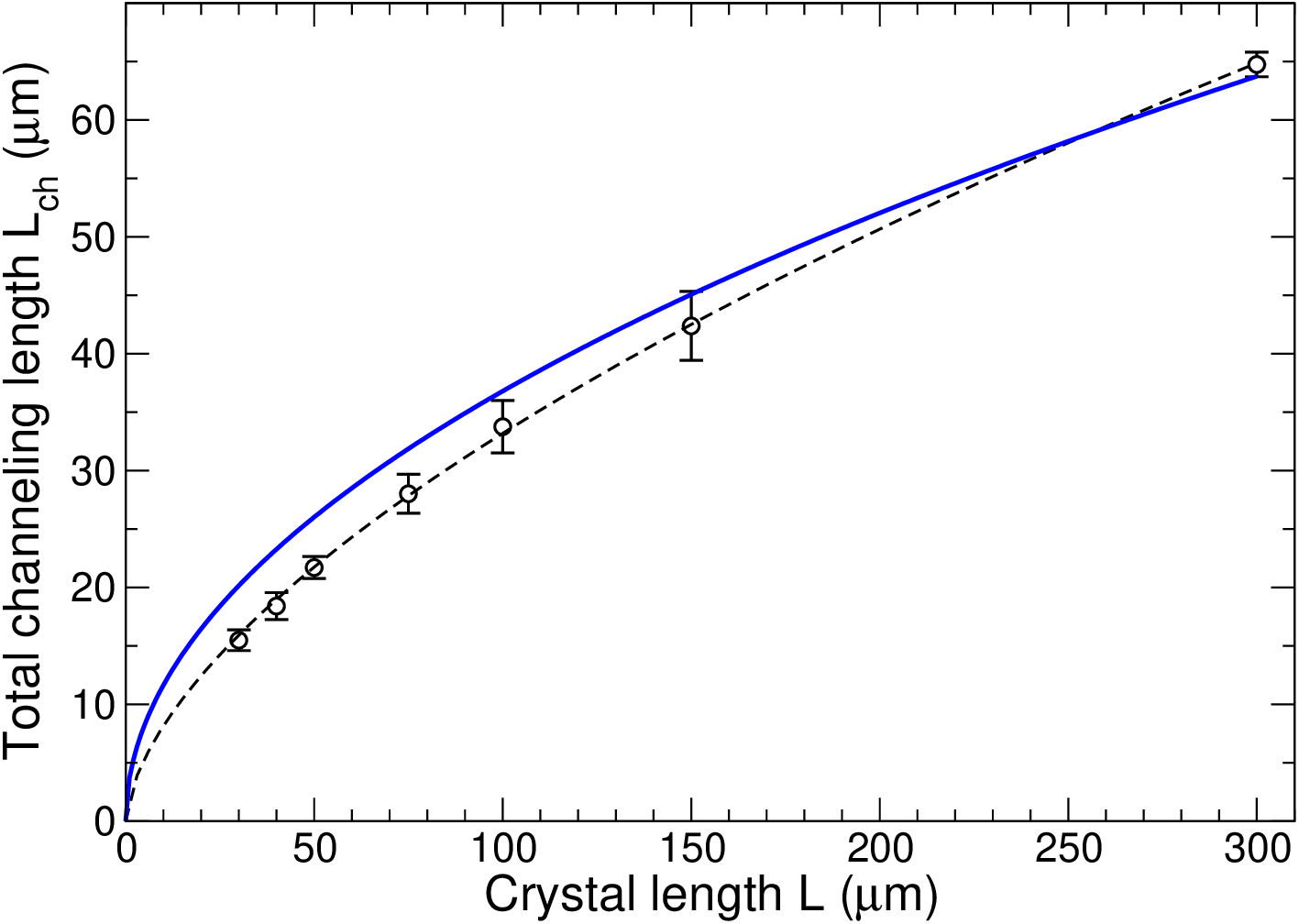}
\caption{Total channeling length $\Lch$ for  855 MeV electrons
in straight Si(110) channel versus crystal length.
Open circles stand for the calculated values, solid and dashed curves
present the dependencies $\Lch\propto L^{0.5}$ and $\Lch\propto L^{0.61}$,
respectively.
See also explanations in the text.}
\label{Figure_04.fig}
\end{figure}

Let us introduce another characteristic length, the total channeling length $\Lch$, 
which can be used to quantify the channeling process.
This quantity is defined as an average length of all channeling 
segments per trajectory.
Due to the rechanneling process, 
the total channeling length $\Lch$ is an increasing function of the crystal length $L$.
Under the assumption that the probability of rechanneling does not depend on 
the penetration length one can expect that 
$\Lch \propto \sqrt{L}$ for $L\gg L_{\rm p1},L_{\rm p2}$.
In this limit the number of rechanneling events is large enough to justify the
diffusion approach in estimating $\Lch$.
The calculated values of $\Lch$ (open circles) versus $L$ for 
855 MeV electrons in straight Si(110) along with the fitting curves are 
presented in Figure \ref{Figure_04.fig}.
The solid line stands for the dependence $\Lch = 3.68 \sqrt{L}$ which seems to be adequate 
for large values of $L$ but overestimates $\Lch$ in the range $L<100$ micron. 
The dashed line stands for the best fit which is described by the formula
$\Lch = 2L^{0.61}$.  

%%%%%%%%%%%%%%%%%

%%%%%%%%%%%%%%%%%%%%%%%%%%%
\section{Concluding remarks}

Using the newly developed code \cite{NewPaper_2013}, 
which was implemented as a module in the MBN Explorer package 
\cite{MBN_ExplorerPaper},
we have performed the Monte Carlo simulations of trajectories of  
ultra-relativistic electrons in oriented (110) straight and bent single 
Si crystal. 
The description of the particle motion is given in classical terms by 
solving the relativistic equations of motion which account for the 
interaction between the projectile and the crystal atoms. 
The probabilistic element is due a random choice 
of transverse coordinates and velocities of the projectile at the 
crystal entrance as well as by accounting for the random positions of 
the crystal atoms due to thermal vibrations. 

The calculations were carried out for several values of the bending radius $R$ and
crystal length $L$.
The chosen energies  195, 600 and 855 MeV of electrons are of interest in connection with 
the ongoing experiments with electron beams at Mainz Microtron 
\cite{Backe_EtAl_2011,BackeLauth_2012}.

The simulated trajectories are used as the input data for numerical analysis 
of channeling conditions and properties.
The calculated data include the channel acceptance as a function of 
the bending parameter $C$ and of electron energy $\E$;
the number of primarily channeled particles $N_{\rm ch 0}$ and the number of 
particles in the channeling mode $\Nch$ as functions of the penetration distance $z$.
By comparing the dependencies $N_{\rm ch 0}(z)$ and  $N_{\rm ch 0}(z)$ 
we have demonstrated the important role of rechanneling for projectile electrons.

By analyzing the simulated trajectories we estimated of
the electron dechanneling lengths for various $\E$ and $C$. 
The obtained results, which are summarized in Table \ref{Table_lengths},
are more accurate than those calculated recently 
\cite{KKSG_simulation_straight,BentSi110_2011} since the latter data 
were based on the erroneous model of ultra-relativistic electron--atom elastic 
scattering.
We have demonstrated that the model overestimates the mean scattering angle,

Finally, we have performed quantitative analysis of the total length $\Lch$ of 
all channeling segments per a simulated trajectory. 
This quantity can be measured experimentally by analyzing the radiation emitted within 
a large cone, $\theta \gg \gamma^{-1}$, as a function of the crystal length.
As a prime further step in application of the simulated trajectories 
we plan to perform the calculations of the emission spectra in a broad range of the
photon energies. 
This will allow us to carry out comparison with the available experimental data and
to analyze the influence of the crystal bending on the spectra.
Similar calculations are to be performed for the case of positron channeling.

Another important set of calculations concerns 
simulation and analysis of the channeling process and computation of the 
emission spectra from sub-GeV electrons and positrons in crystalline undulators.
The results of this work, which is currently in progress, will be published
elsewhere.

%%%%%%%%%%%% Acknowledgments
\ack
We are grateful to Hartmut Backe and Werner Lauth for fruitful and
stimulating discussions. 
The work was supported by the European Commission CUTE 
(the CUTE-IRSES project, grant GA-2010-269131). 
The possibility to perform complex computer 
simulations at the Frankfurt Center for Scientific Computing is 
gratefully acknowledged.

%%%%%%%%%%%%%%%%%%%%%%%%%%%%%%%%%%%%%%%%%%%%%%%%%%%%%%%

%%%%%%%%%%%%%%%%%%%%%%%%%%%%%%%%%%%%%%%%%%%%%%%%%%%%%%%
%%%%%%%%%%%%%%%%%%%%%%%%%%%%%%%%%%%%%%%%%%%%%%%%%%%%%%%
%%%%%%%%%%%%%%%%%%%%%%%%%%%%%%%%%%%%%%%%%%%%%%%%%%%%%%%
\end{document}